\input{aipcheck}

\documentclass[
    ,final            
  ]
  {aipproc}

\usepackage{amsmath,graphicx}

\newcommand{\true}{{\text{true}}}
\newcommand{\obs}{{\text{obs}}}
\newcommand{\trueobs}{{\text{true,obs}}}

\layoutstyle{8x11double}

\begin{document}

\title{More on Non-standard Interaction at MINOS}


\author{Hiroaki Sugiyama}{
  address={SISSA, Via Beirut 2-4, 34014 Trieste, Italy}
}



\begin{abstract}
 We discuss about effects of the non-standard interaction
of neutrinos with matter on the $\nu_e$ appearance search
in the MINOS experiment.
 We consider the effects of the complex phase of the interaction
and of the uncertainty on $\theta_{23}$ also.
 We show that the oscillation probability can be so large
that can not be explained by the ordinary oscillation.
 We show also how much constraints on the non-standard effects
can be improved
if the experiment does not observe $\nu_e$ appearance signal.
\end{abstract}

\maketitle

\section{Introduction}

 Oscillations of three generation neutrinos are parametrized
by two mass squared differences, three mixing angles, and
one CP violating phase.
 Among these parameters,
a tiny mixing angle $\theta_{13}$ and the phase $\delta$
have not been measured yet.
 Measuring tiny effects of $\theta_{13}$ and $\delta$
is the main purpose of future oscillation experiments.
 On the other hand,
such precision measurements will be sensitive
to effects of new physics also.
 In this talk, we investigate possibilities to see
new physics effects on the $\nu_e$ appearance search
with $\nu_\mu $ beam in the MINOS experiment%
~\cite{Michael:2006rx}.
 This talk is based on \cite{Kitazawa:2006iq}.

 We consider the non-standard interaction (NSI) of neutrinos
with matter as an example of new physics.
 The interaction is introduced in a model-independent way
by the following effective Lagrangian
\begin{eqnarray}
 {\cal L}_{\text{eff}}
 = \epsilon_{\alpha\beta}^{fP} G_{\text{F}}
   (\overline{\nu_\alpha}\gamma_\mu P_L \nu_\beta)
   (\overline{f} \gamma^\mu P f ),
\label{NSI}
\end{eqnarray}
where $\alpha$ ($= e, \mu, \tau$) and $\beta$ stand for flavors,
$G_{\text{F}}$ is the Fermi coupling constant,
$P$ ($=P_L, P_R$) denotes the projection operator onto
the left-handed one or the right-handed one,
and $f$ ($= e, u, d$) represents the fermions existing in matter.
 Then,
the Hamiltonian
in the flavor basis is modified from the standard one
and it is given by
\begin{eqnarray}
&&\hspace*{-10mm}
H
= \frac{1}{2E}
U_{\text{MNS}}
\left(
\begin{array}{ccc}
0 & 0 & 0\\
0 & \Delta m^2_{21} & 0\\
0 & 0 & \Delta m^2_{31}
\end{array}
\right)
U_{\text{MNS}}^\dagger\nonumber\\
&&\hspace*{10mm}
{}+A\left(
\begin{array}{rrr}
1+ \epsilon_{ee} & \epsilon_{e\mu} & \epsilon_{e\tau}\\[2mm]
\epsilon_{e\mu}^\ast & \epsilon_{\mu\mu} & \epsilon_{\mu\tau}\\[2mm]
\epsilon_{e\tau}^\ast & \epsilon_{\mu\tau}^\ast & \epsilon_{\tau\tau}
\end{array}
\right),
\label{matterV}
\end{eqnarray}
where $A \equiv \sqrt{2} G_{\text{F}} n_e$
with the electron number density $n_e$,
and $\epsilon_{\alpha\beta} \equiv \sum_P (\epsilon_{\alpha\beta}^{eP}
+ 3 \epsilon_{\alpha\beta}^{uP} + 3 \epsilon_{\alpha\beta}^{dP})$.
 The first term of the right hand-side of (\ref{matterV})
is for the oscillation in vacuum
and the second term is the matter potential with NSI\@.
 In this talk,
a typical matter density
$\rho = 2.7\text{g}\cdot\text{cm}^{-3}$ is used
and then we have $A \simeq 10^{-13}\text{eV}$.
 On the other hand,
the size of the vacuum part of the Hamiltonian
are controlled by $\Delta m^2/2E$, and it is $\sim 10^{-13}\text{eV}$
for the MINOS experiment.
 Thus,
we expect some NSI effect in the MINOS experiment
if $\epsilon_{\alpha\beta}$ can be $\sim 1$.

\begin{figure}[t]
\begin{minipage}{80mm}
\includegraphics[origin=c, angle=-90, scale=0.35]{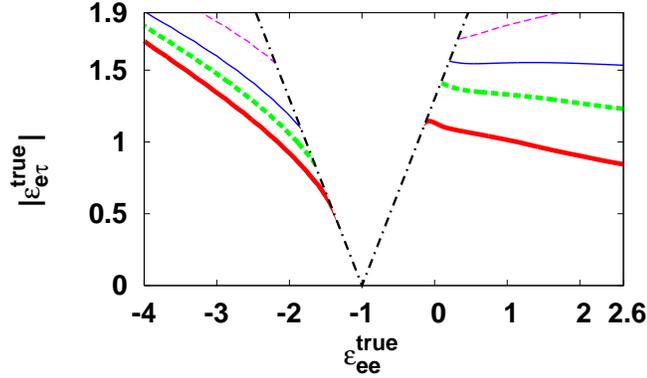}\\[-13mm]
\caption{
 Thin dashed, thin solid, bold dashed, and bold solid curves
are obtained for $\sin^2{2\theta_{13}^\true} = 0, 0.05, 0.1$, and 0.16,
respectively.
 If true values exist above these curves, the MINOS experiment
can see the NSI effect.
 The region above the dash-dotted line has been excluded
by the atmospheric neutrino measurement and the K2K experiment.
}
\label{fig:sens}
\end{minipage}
\end{figure}
\section{Constraints on $\epsilon_{\alpha\beta}$}

 Experiments involving charged leptons
strongly bound the effective interaction ($\ref{NSI}$)
if $SU(2)_L$ symmetry is imposed.
 From a model independent point of view, however,
the strength $\epsilon_{\alpha\beta}^{fP}$
of the interaction ($\ref{NSI}$)
can be independent of the strength
of the charged lepton version of ($\ref{NSI}$)
because we know $SU(2)_L$ is broken.
 Therefore,
$\epsilon_{\alpha\beta}^{fP}$ can be constrained only
by experiments with neutrinos.~\footnote{Actually,
constraints on $\epsilon_{\alpha\mu}^{fP}$ are
obtained by experiments with muon through loop effects.}
 Using the results obtained in \cite{Davidson:2003ha},
we have constraints on the elements of (\ref{matterV}).
We put
$|\epsilon_{e\mu}|=\epsilon_{\mu\mu}=|\epsilon_{\mu\tau}|=0$
because they are constrained enough.
 The constraints on $\epsilon_{ee}$ and $|\epsilon_{e\tau}|$ are
\begin{eqnarray}
 -4 < \epsilon_{ee} < 2.6, \ \ \ \
 |\epsilon_{e\tau}| < 1.9.
\end{eqnarray}
 The constraint on $\epsilon_{\tau\tau}$ in \cite{Davidson:2003ha}
is rather loose but it is improved by
the atmospheric neutrino measurement and the K2K experiment%
~\cite{Friedland:2005vy}.
 In order to reproduce the observed $\nu_\mu$ disappearance
for $\Delta m^2_{21} = 0$ and $\theta_{13} = 0$, which is
 $1-P(\nu_\mu \to \nu_\mu)
 = \sin^2{2\theta_{23}^{\text{obs}}}
   \sin^2((\Delta m^2_{31})^{\text{obs}} L/4E)$,
the parameters in vacuum ($\theta_{23}$, $\Delta m^2_{31}$) and non-standard interactions should satisfy
\begin{eqnarray}
&&\hspace*{-10mm}
 \epsilon_{\tau\tau} = \frac{|\epsilon_{e\tau}|^2}{1+\epsilon_{ee}},
\label{atmK2K}\\[1mm]
&&\hspace*{-10mm}
\cos{2\theta_{23}}\nonumber\\[-1mm]
&&\hspace*{-5mm}
= \frac{
s_\beta^2(1+c_\beta^2)
+ 4 c_\beta^2 \cot{2\theta_{23}^{\text{obs}}} \sqrt{ (1+\cot^2\!{2\theta_{23}^{\text{obs}}}) }
}{ (1+c_\beta^2)^2 + 4\,c_\beta^2\,\cot^2\!{ 2\theta_{23}^{\text{obs}} } },\\
&&\hspace*{-10mm}
\Delta m^2_{31}
= \frac{2 \left( \Delta m^2_{31} \right)^{\text{obs}} }
{ \sqrt{ \{ (1+c_\beta^2) \cos{2\theta_{23}} - s_\beta^2 \}^2
 + 4 c_\beta^2 \sin^2\!{2\theta_{23}} } },
\end{eqnarray}
where $\tan\beta \equiv |\epsilon_{e\tau}|/(1+\epsilon_{ee})$,
$c_\beta$ and $s_\beta$ denote $\cos\beta$ and $\sin\beta$,
respectively.
 Furthermore,
in order to be consistent with the sub-GeV data of atmospheric $\nu_\mu$,
where the matter effect is suppressed,
the parameters in vacuum should satisfy
\begin{eqnarray}
 \cos{2\theta_{23}} < 0.5, \ \ \ \
 |\Delta m^2_{31}| < 5\times 10^{-3} \text{eV}^2.
\label{subGeV}
\end{eqnarray}
 The conditions give an upper-bound on $|\tan\beta|$,
and then we can constrain $\epsilon_{\tau\tau}$ also
as $|\epsilon_{\tau\tau}| < |\tan\beta|^{\text{max}}
 |\epsilon_{e\tau}|^{\text{max}}$.
 We use the conditions (\ref{atmK2K})-(\ref{subGeV})
for also the case with $\Delta m^2_{21}\neq 0$ and $\theta_{13}\neq 0$
for simplicity.

\section{Analyses and Results}
 In our analysis,
we calculate numbers of $\nu_e$ events (signal and background)
for 13 bins of 0.5GeV width
within 1-7.5GeV of the reconstructed energy.
 We assume $16\times 10^{20}$POT which corresponds to
about 5 years of running.
 Systematic errors are ignored for simplicity.
 Two parameters are fixed as
$\sin^2{2\theta_{12}} = 0.8$,
$\Delta m^2_{21} = 8\times 10^{-5}\text{eV}^2$
throughout this talk.

\begin{figure}[t]
\begin{minipage}{80mm}
\includegraphics[origin=c, angle=-90, scale=0.35]{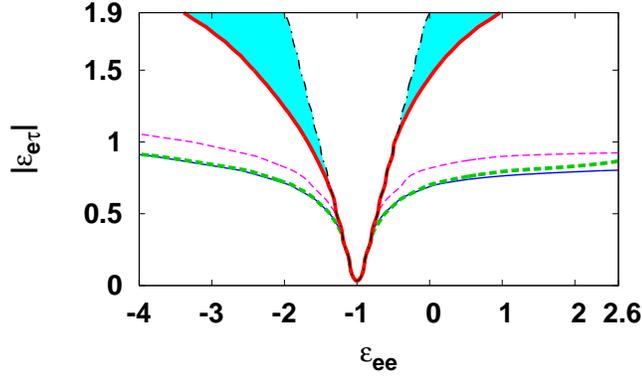}\\[-13mm]
\caption{
 The results for the case without $\nu_e$ appearance signal are shown.
 The region above the dash-dotted line has been excluded
by the atmospheric neutrino measurement and the K2K experiment.
 Additionally to that,
the gray region can be excluded by the MINOS experiment.
}
\label{fig:bound}
\end{minipage}
\end{figure}
 First,
we investigate the possibility to see the effect of NSI
in the MINOS experiment.
 ``Data'' are generated with NSI
and we try to fit the ``data'' without NSI.
 If the $\Delta \chi^2$ that corresponds to the fitting is larger than 4.6,
it means that the MINOS experiment can exclude\
the case with $\epsilon_{ee} = |\epsilon_{e\tau}| = 0$
and see the NSI effect at 90\%CL\@.%
~\footnote{
 Since pure information about $\theta_{13}$ can be obtained
in the future reactor experiments,
we try to extract information about
$\epsilon_{ee}$ and $|\epsilon_{e\tau}|$ only.
 Therefore, we rely on the analysis for 2 degrees of freedom
in which $\Delta \chi^2 = 4.6$ corresponds to 90\%CL\@.
 If we try to extract information about $\theta_{13}$ also,
we should use the analysis for 3 degrees of freedom
and then the results get worse with $\Delta \chi^2 = 6.3$.}
 For the generation of the ``data'' with NSI,
we use
\begin{eqnarray}
&&\hspace*{-13mm}
 \delta^\true = \text{arg}(\epsilon_{e\tau}^\true) = 0,\\
&&\hspace*{-13mm}
 s_{23}^\trueobs = 1/\sqrt{2}, \ \ \
 (\Delta m^2_{31})^\trueobs = 2.7\times 10^{-3}\text{eV}^2.
\label{obs}
\end{eqnarray}
 We fix $\sin^2{2\theta_{13}^\true}$ also
but take several values for that.
 On the other hand,
the values of parameters for fitting without NSI are
\begin{eqnarray}
&&\hspace*{-13mm}
 0 \leq \sin^2{2\theta_{13}} \leq 0.16,\\
&&\hspace*{-13mm}
 \delta = -\pi/2, \ \ \
 s_{23} = 0.8 \ (\sin^2{2\theta_{23}}=0.92),
\label{max}\\
&&\hspace*{-13mm}
 \Delta m^2_{31} = 2.7\times 10^{-3}\text{eV}^2.
\end{eqnarray}
 $\Delta \chi^2$ is minimized with respect to $\theta_{13}$
within the region.
 Note that the choice of (\ref{max}) is
for large $P(\nu_\mu \to \nu_e)$ in the standard oscillation,
and hence it is a pessimistic one for the search of the NSI effect.
 The result is shown in Fig.~\ref{fig:sens}.
 The region above the dash-dotted line has been excluded by
$|\tan\beta| < 1.3$ obtained with (\ref{subGeV}) and (\ref{obs}).
 The thin dashed, thin solid, bold dashed, and bold solid curves
are results for $\sin^2{2\theta_{13}^\true} = 0, 0.05, 0.1$, and 0.16,
respectively;
 If nature chooses true values above the curves,
the number of $\nu_e$ appearance signal becomes so large
that cannot be explain with $\theta_{13}$ only.
 Then,
it is possible to see the NSI effect in the MINOS experiment.

 Next,
let us consider the case without $\nu_e$ appearance signal.
 In this case,
bounds on $\epsilon$ will be obtained by the search.
 ``Data'' for the analysis is generated by
\begin{eqnarray}
&&\hspace*{-10mm}
 \theta_{13}^\true = \epsilon_{ee}^\true = |\epsilon_{e\tau}^\true| = 0,\\
&&\hspace*{-10mm}
 s_{23}^\true = 1/\sqrt{2}, \ \ \
 (\Delta m^2_{31})^\true = 2.7\times 10^{-3}\text{eV}^2,
\end{eqnarray}
and we try to fit the ``data'' by using NSI\@.
 If the fitting is failed,
we can exclude the values of $\epsilon$.
 For the fitting procedure,
we use
\begin{eqnarray}
&&\hspace*{-10mm}
 0 \leq \sin^2{2\theta_{13}} \leq 0.16,\\
&&\hspace*{-10mm}
 \delta = 0, \ \ \
 \text{arg}(\epsilon_{e\tau}) = 0, \pi/2, \pi, -\pi/2,\\
&&\hspace*{-10mm}
 s_{23}^\obs = 0.6, 1/\sqrt{2}, 0.8,\\
&&\hspace*{-10mm}
 (\delta m^2_{31})^\obs = 2.7\times 10^{-3}\text{eV}^2.
\end{eqnarray}
 Note that phases appear as $\delta + \text{arg}(\epsilon_{e\tau})$
approximately, which is exact for $\Delta m^2_{21} = 0$,
%
and then we can put $\delta=0$.
 $\Delta \chi^2$ for this analysis is minimized
with respect to $\theta_{13}$ and three values of $s_{23}^\obs$.
 The smallest and the largest values of $s_{23}^\obs$ are
obtained from $\sin^2{2\theta_{23}^\obs} = 0.92$.
 In Fig.~\ref{fig:bound},
results for $\text{arg}(\epsilon_{e\tau}) = 0, \pi/2, \pi$, and $-\pi/2$
are shown by thin dashed, thin solid, bold dashed, and bold solid curves.
 The region bellow the curves can be consistent
with the case of no $\nu_e$ appearance at 90\%CL,
and then we should take the bold solid curve as a pessimistic choice
for the exclusion of $\epsilon$ in the MINOS experiment.
 The dash-dotted line is given by (\ref{subGeV})
and the region above the line has been excluded
by atmospheric neutrino measurement and the K2K experiment.
 Hence,
the gray region shows the possible improvement
of the bound on $\epsilon$ in the MINOS experiment.

\section{Conclusions}
 We have investigated possibilities to obtain information
about NSI effect with the search of $\nu_\mu \to \nu_e$ oscillation
in the MINOS experiment.
 We have shown that it is possible to find the effect in the experiment.
 Even if the experiment does not find $\nu_e$ appearance signal,
some part of $\epsilon_{ee}$-$\epsilon_{e\tau}$ space
can be excluded by the result.
 Therefore,
ongoing and future oscillation experiments are very interesting
not only for the precise determination of the mixing parameter
in the lepton sector but also for the new physics search.

\end{document}